\documentclass[11pt]{article}

\usepackage{psfig}
\usepackage[dvips]{epsfig}
\usepackage{wrapfig}
\usepackage{subfigure}
\usepackage{mathptmx}
\usepackage{latexsym}     
\usepackage{amssymb} 
\usepackage{graphicx}
\usepackage[figuresright]{rotating}

\newcommand{\be}{\begin{equation}} 
\newcommand{\ee}{\end{equation}}   
\newcommand{\bea}{\begin{eqnarray}}
\newcommand{\eea}{\end{eqnarray}}
\newcommand{\bean}{\begin{eqnarray*}}
\newcommand{\eean}{\end{eqnarray*}}  
\newcommand{\bef}{\begin{figure}}  
\newcommand{\eef}{\end{figure}}  
\newcommand{\bet}{\begin{table}}
\newcommand{\eet}{\end{table}}
\long\def\hidestart#1\hideend{}

\begin{document}

\title{\bf{More on the continuum limit of gauge-fixed compact $U(1)$ lattice
gauge theory }}
\author
{Asit K. De$^a$\thanks{email: asitk.de@saha.ac.in} and
Tilak Sinha$^a$\thanks{email: tilak.sinha@saha.ac.in} \\
\\
$^a$Theory Group, Saha Institute of Nuclear Physics,\\
1/AF, Salt Lake, Calcutta 700064, India}
\date{}
\maketitle

\begin{abstract}
We have verified various proposals that were suggested in our last paper
concerning the continuum limit of a compact formulation of the lattice
U(1) pure gauge theory in 4 dimensions using a nonperturbative gauge-fixed
regularization. Our study reveals that most of the speculations are largely
correct. We find clear evidence of a continuous phase transition in
the pure gauge theory at "arbitrarily" large couplings. When probed with
quenched staggered fermions with U(1) charge, the theory clearly has a 
chiral transition for large gauge couplings whose intersection with the
phase transition in the pure gauge theory continues to be a promising
area for nonperturbative physics. We probe the nature of the continuous
phase transition by looking at gauge field propagators in the momentum space
and locate the region on the critical manifold where free photons 
can be recovered.
\end{abstract}

\vspace{0.15cm}

$~~~~$PACS: 11.15.Ha; 12.20.Ds
\vspace{0.11 cm}

\section*{Introduction}

The existence and nature of the possible continuum limits of a U(1) theory
formulated on the lattice
have been a long standing issue in Lattice field theory. QED is arguably
the most phenomenologically successful quantum field theory.
Nevertheless its behaviour at short distance is still under speculation. 
The indication from perturbation theory is that, to evade Landau poles QED
becomes {\em trivial} as one tries to remove the cutoff and interpret it as a
fundamental theory (as opposed to an effective theory valid upto some
cutoff). The interest in this question is fundamental rather than
phenomenological since in QED the landau pole lies beyond the plank scale. 
The motivation in its resolution stems from the belief that all
non-asymptotically free theories are trivial and therefore cannot be 
fundamental. The existence of QED as a fundamental theory could have a
profound impact on the unification schemes \cite {unif}.

The search for a nontrivial continuum limit for QED on the lattice 
has not yet borne fruit. With the usual compact pure gauge Wilsonian 
action there is a confinement-deconfinement phase transition 
generally understood in terms of monopole condensation \cite{flux,monopole}. 
This transition was believed to be first
order for a long time but the evidence was inconclusive. A couple of years
ago, a high statistics finite size scaling analysis has actually confirmed
that the phase transition is indeed first order \cite{fss}. With the inclusion 
of fermions incidentally, this phase transition coincides 
with a chiral phase transition as well. 

The effect of the inclusion of a four-fermi interaction term to the standard
Wilson action, both with the compact and non-compact formulations is
also being studied for quite some time \cite{azco0, azco1}. 
Recently in a series of papers \cite{ffermi1,ffermi2}
it has been concluded that the inclusion of a four-fermi interaction term 
leads to a continuous chiral phase transition that is distinct from the
coulomb-confinement phase transition beyond a certain value of the
four-fermi coupling and is actually completely controlled by this coupling.
Unfortunately, the study also indicates that the chiral phase 
transition is logarithmically trivial in conformity with traditional
beliefs. An earlier study using the non-compact formulation has  
yielded similar results \cite{ffermi3}.  

There has also been considerable interest in the search for a continuum
limit of QED with nonperturbative properties differing from our familiar
QED accessible through perturbation theory. For a review see \cite{jiri1}.
Such theories, apart from being
interesting by themselves can be useful in the nonperturbative description
quantum field theories beyond the standard model.

It has been remarked before that 
inclusion of fermions to the compact pure Wilson action does not lead to a
new phase transition but if scalars are also included, continuum limits 
with interesting non-perturbative features have been claimed 
to emerge \cite{jiri2}. One of these continuum
limits is associated with a {\em tricritical} point where different critical
manifolds intersect \cite{jiri2}. Signals of interesting new continuum
limits have also been claimed with nonminimal plaquette extensions of the
Wilson action \cite{jiri3,jiri4}.

In a previous study \cite{paper1}, we have
investigated the continuum limit of a compact formulation of the lattice
U(1) gauge theory in 4 dimensions using a novel regularization
that was born in the quest for a lattice formulation of chiral gauge theory
\cite{golter}. This regularization of lattice U(1) theory
was originally devised to tame the 'rough gauge' problem of lattice chiral
gauge theories \cite{reduced}.
Because of the lack of gauge-invariance of the lattice chiral
gauge theories and the gauge-invariant measure,
the longitudinal gauge degrees of freedom ({\em lgdof}) couple
nonperturbatively to the physical degrees of freedom. To decouple the
{\em lgdof} which are radially frozen scalar fields, a nonperturbative gauge
fixing scheme (corresponding to a local renormalizable covariant gauge
fixing in the naive continuum limit) for the compact U(1) gauge fields was
proposed. 
The regularization has a 
bigger parameter space which apart from the standard Wilson term 
includes a gauge-fixing term for compact gauge fields and a
mass-counterterm. A key feature of this gauge fixing scheme is that
the gauge fixing term is not the exact square of the expression used in the
gauge fixing condition and as a result not BRST-invariant (as required by
Neuberger's theorem \cite{neu} for compact gauge fixing). It has, in
addition, appropriate irrelevant terms to make the perturbative vacuum
$U_{\mu x}=1$ unique. Because the gauge fixing term obviously breaks gauge
invariance, one needs to add counter-terms to restore manifest gauge
symmetry.

It has been shown that with the above parameterization, 
for weak gauge couplings there exists
a continuous phase transition 
(that has been called FM-FMD)
at which the {\em lgdof} decouple, and 
the U(1) gauge symmetry is restored \cite{bock1}. This phase transition
has to be accessed from the FM phase since the FMD phase is a phase with
broken rotational symmetry.

Our investigations in \cite{paper1} led to a clear evidence 
for a continuous phase transition 
in the pure gauge theory 
for all values of the 
gauge couplings 
studied  
if the coefficient of the gauge fixing term is
kept adequately {\em large}. We speculated that the theory will sustain this 
feature irrespective of how large the gauge coupling is. On the other hand
we found that the phase transition is first order if the coefficient of the
gauge fixing term {\em small}. We naturally assumed that there exists
a critical value for the coefficient of the gauge fixing term
corresponding to a given gauge coupling, below which the phase transition is 
first order and is continuous above it. We realized that for a range of 
gauge couplings this result would be the evidence for the existence 
of a multicritical line on the FM-FMD interface.

When probed with quenched staggered fermions with U(1)
charge, the theory also displayed an unambiguous signal of a 
chiral transition for large gauge
couplings. It seemed to be a likely possibility that the line where 
the  chiral phase transition intersects the FM-FMD transition 
coincides with the multicritical line where the order of the 
FM-FMD transition changes. In such an event this multicritical line 
could become a strong candidate for non-perturbative physics.

In this paper, we have embarked on the task of investigating the truth of
the various speculative propositions that were put forth in our exploratory 
paper \cite{paper1}.
 
We have confirmed our speculation that indeed if the gauge 
coupling is increased arbitrarily (numerically, to very large values) 
the continuous FM-FMD transition is retrieved by adequately increasing the 
coefficient of the gauge fixing term.

We have also determined the location of the 
multicritical line precisely as well as the
the region where the chiral phase transition meets the
FM-FMD transition and we have verified how far 
they coincide to our precision.

Further in this paper, we have made an effort to probe into the 
nature of the continuous FM-FMD phase transition by looking at 
the gauge field propagator in momentum space to see where (if anywhere)
in our critical manifold, free photons can be extracted. 

The results of our study promises to nurture both the interests, 
the triviality issue and the possibility of new continuum limits,
in lattice $U(1)$ theories.

\section*{The Regularization}

\begin{figure}[!t]
\centering
\subfigure{
\includegraphics[width=2in,clip]{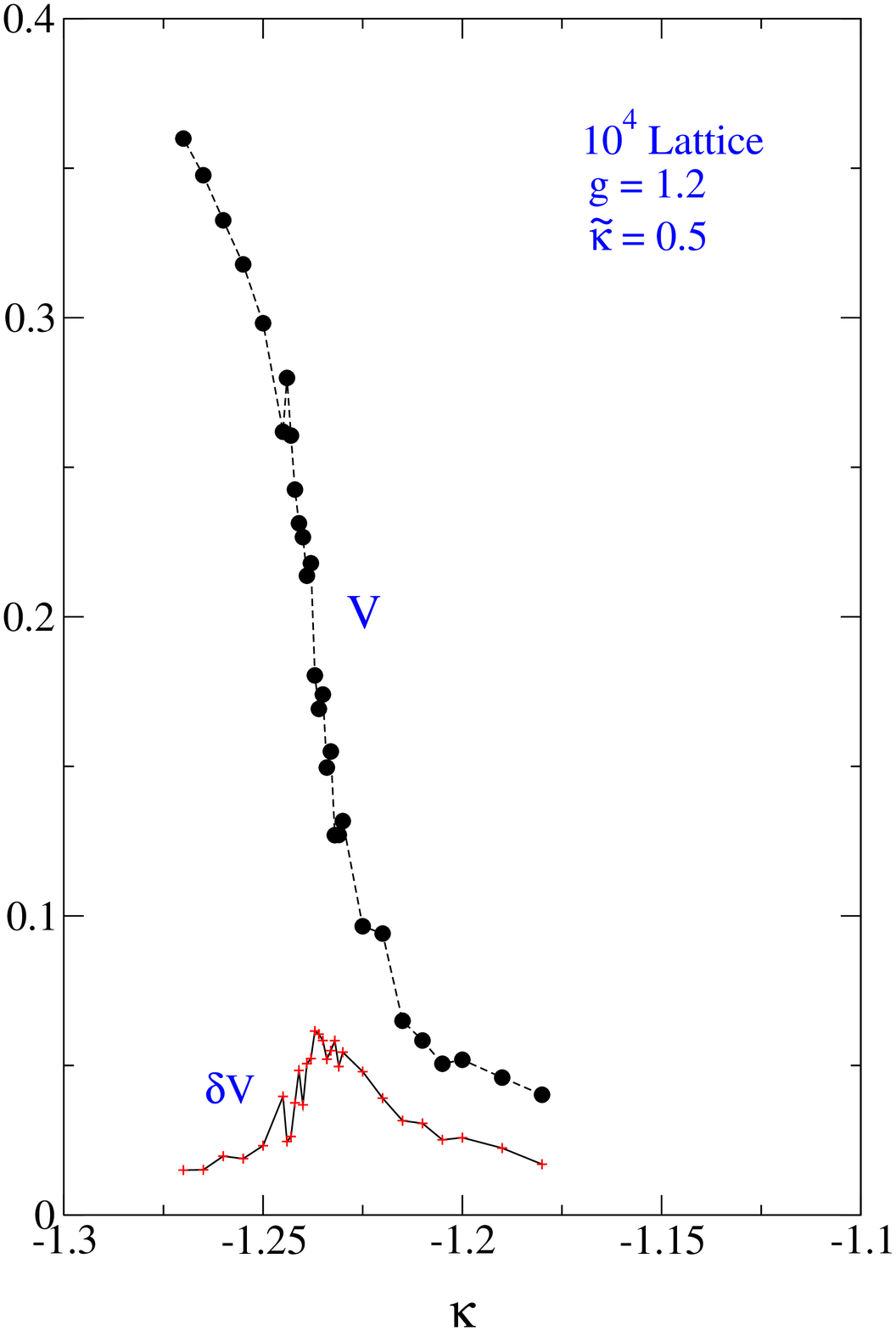}}
\subfigure{
\includegraphics[width=2in,clip]{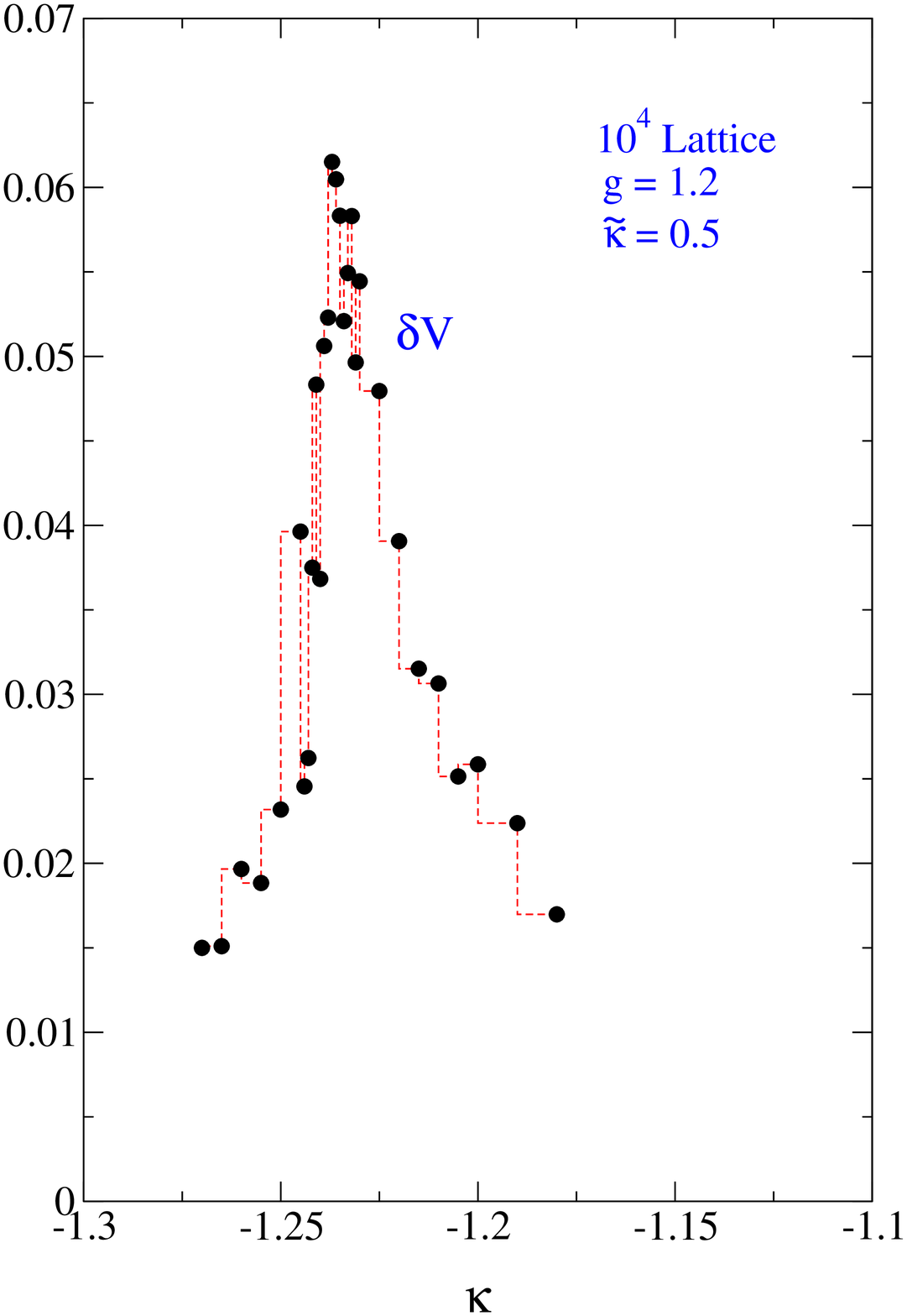}}
\caption{The order parameter $V$ (left) and its fluctuation $\delta V$
(right) as a function of $\kappa$ across the critical point in a region
of the FM-FMD interface where the phase transition is continuous. The
figure shows the efficiency of the $\delta V$ plot in locating the critical
point precisely.}
\label{vndelv}
\end{figure}

\begin{figure}[!t]
\centering
\includegraphics[width=3in,clip]{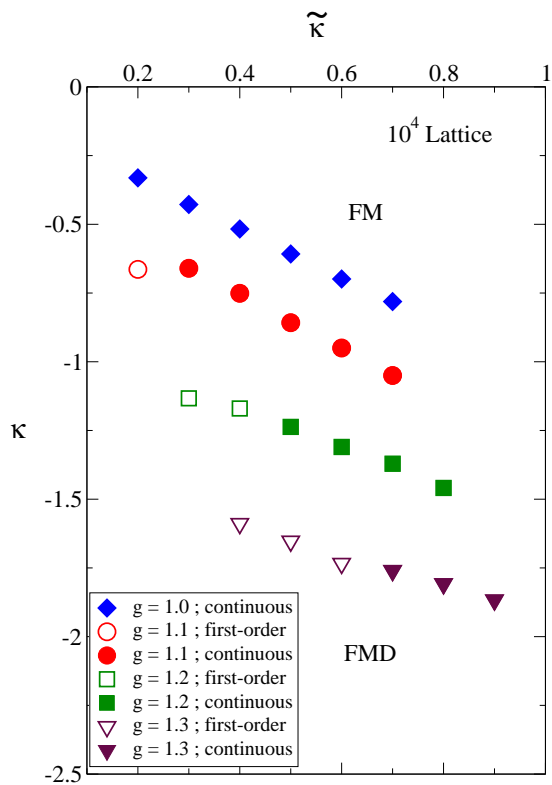}
\caption{The carefully redetermined FM-FMD phase transition lines in the
$\kappa-\tilde{\kappa}$ plane for four different gauge couplings. Hollow
symbols represent first-order phase transition and filled symbols represent
continuous phase transition. The tricritical line divides the region of
hollow symbols from that of filled symbols.}
\label{nuphase}
\end{figure}

The action for the compact gauge-fixed U(1) theory 
\cite{golter}, where the ghosts are free and decoupled, is:
\be
S[U] = S_g[U] + S_{gf}[U] + S_{ct}[U]. 
\label{action}
\ee

$S_g$ is the usual Wilson plaquette action,
\be
S_G = \frac{1}{g^2} \sum_{x\,\mu <\nu} \left( 1 -
{\rm Re}\, U_{\mu\nu x} \right) \label{plact}
\ee
where $g$ is the gauge coupling and $U_{\mu x}$ is the group valued U(1)
gauge field. 

$S_{gf}$ is the BRST-noninvariant compact gauge fixing term, 
\be
S_{gf} = \tilde{\kappa} \left( \sum_{xyz} \Box(U)_{xy}
\Box (U)_{yz} - \sum_x B_x^2 \right) \label{gfact}
\ee
where $\tilde{\kappa}$ is the coefficient of the gauge fixing term, 
$\Box(U)$ is the covariant lattice Laplacian and
\be
B_x  = \sum_\mu \left( \frac{{\cal A}_{\mu, x-\mu} +
{\cal A}_{\mu x}}{2}\right)^2,
\ee
where ${\cal A}_{\mu x} {\rm = Im}\, U_{\mu x}$. $S_{gf}$ is not
just a naive transcription of the continuum covariant gauge fixing  
term, it has in addition appropriate irrelevant terms. This makes the action 
have an unique absolute minimum at $U_{\mu x}= 1$,
validating weak coupling perturbation theory around $g=0$ or
$\tilde{\kappa}=\infty$ and in the naive continuum limit reduces
to $1/2\xi \int d^4x (\partial_\mu A_\mu)^2$ with $\xi =
1/(2{\tilde{\kappa}}g^2)$. 

Validity of weak  
coupling perturbation theory together with perturbative
renormalizability helps to determine the form of the counter
terms to be present in $S_{ct}$. It turns out that the most
important gauge counterterm is the dimension-two counterterm, namely the 
gauge field mass counterterm
given by,
\be
S_{ct} = - \kappa \sum_{\mu x} \left( U_{\mu x} +
U_{\mu x}^\dagger \right). 
\ee

In the pure bosonic theory there are possible marginal counter-terms including
derivatives. However, in the investigation of the gauge-fixed theory as
given, the dimension-two counterterm has been mostly considered, because it
alone could lead to a continuous phase transition that recovers the gauge
symmetry. It was argued that the marginal counter-terms would not possibly
create new universality classes for the continuum theory corresponding to
large $\tilde{\kappa}$ (for a discussion on other counter-terms, please see
\cite{golter,bock1}).

\section*{Numerical Simulation and Results}

\subsection*{Confirming our Speculations}

\begin{figure}[!p]
\centering
\subfigure{
\includegraphics[width=2.0in,clip]{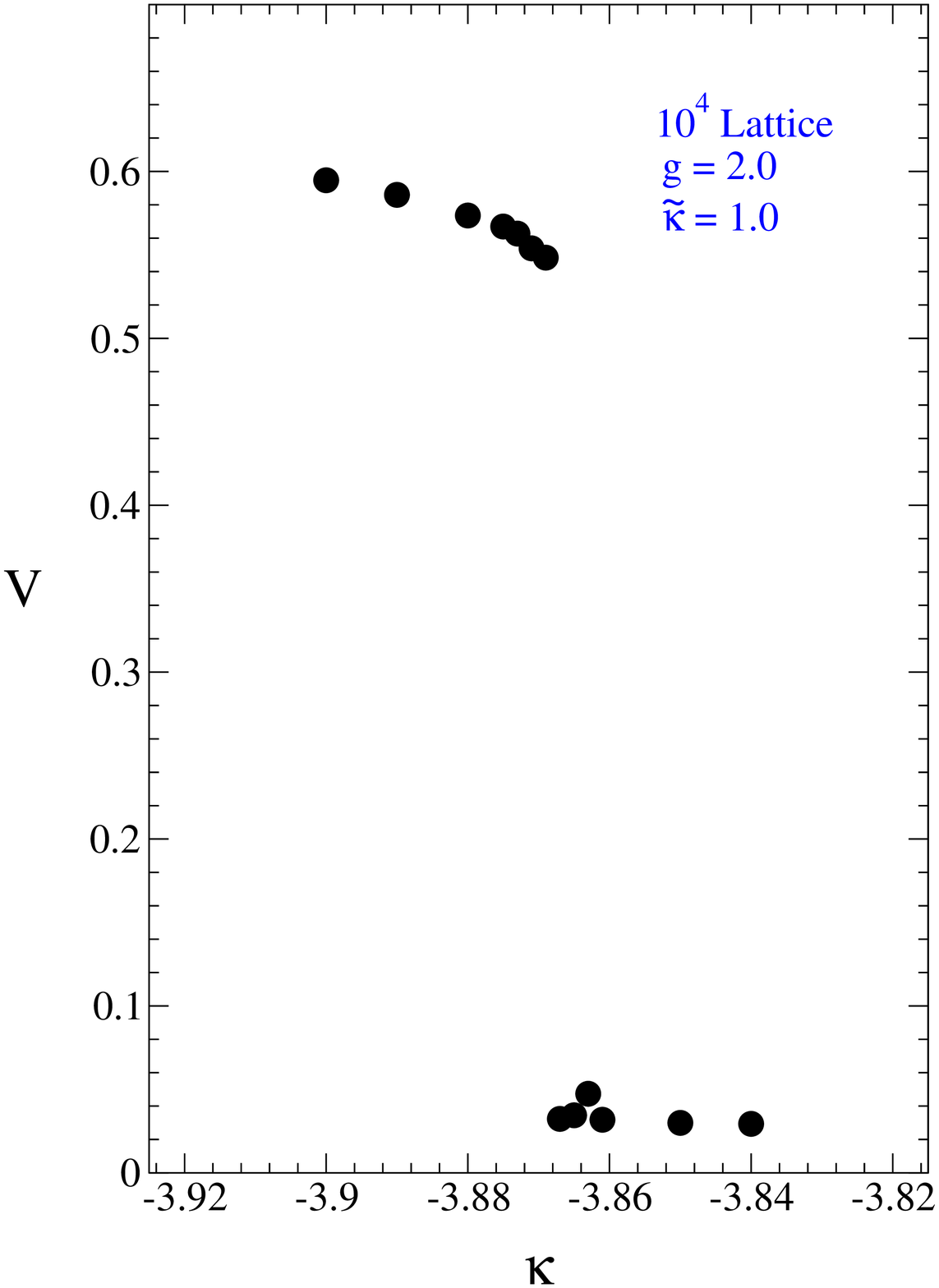}}
\subfigure{
\includegraphics[width=2.0in,clip]{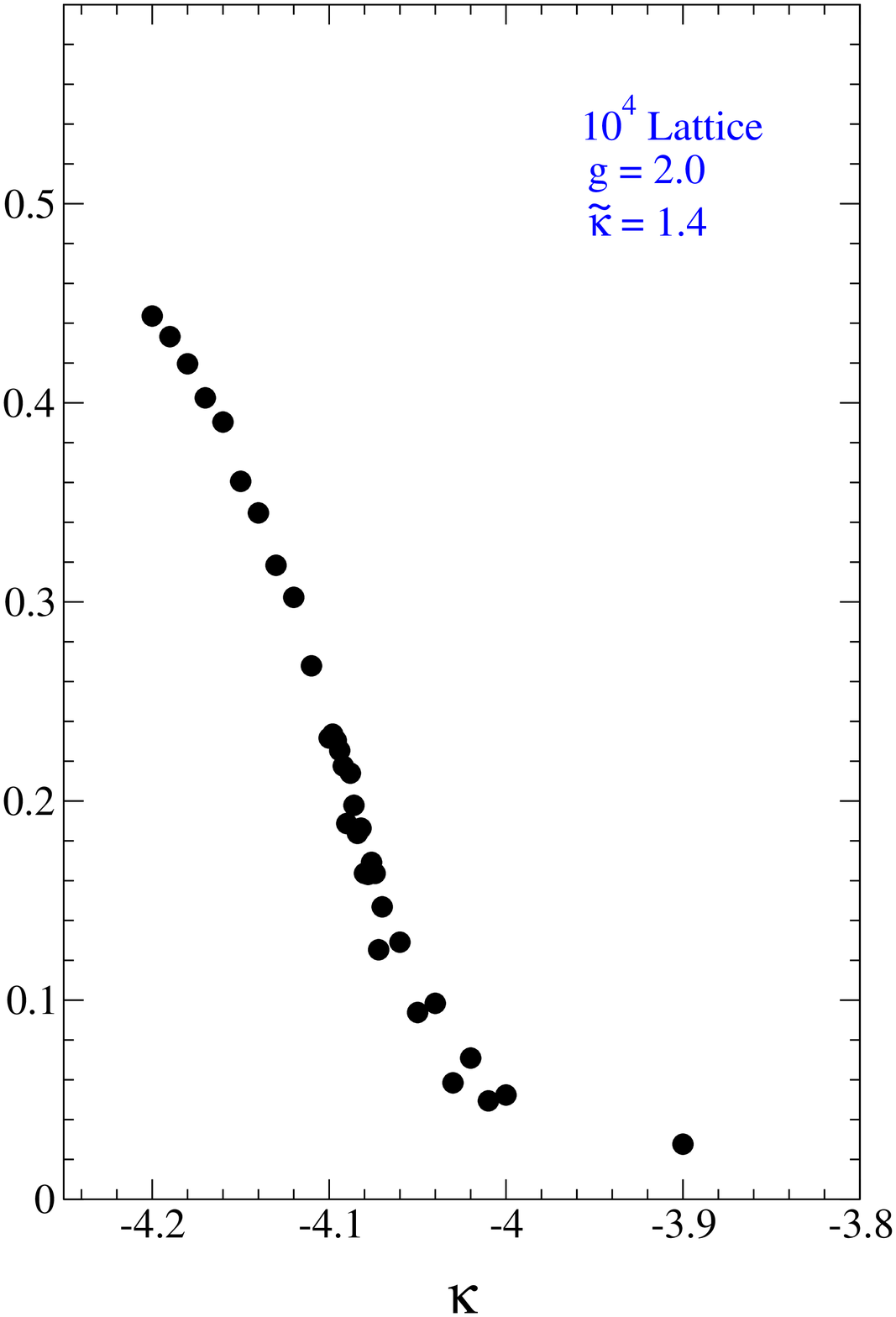}}
\caption{$V$ as a function of $\kappa$
across the critical point at a large value of the gauge coupling
(left) in a region of the FM-FMD interface where the phase transition
is first-order ( $\kappa_{critical} = -3.868$ ) and (right) in a region 
of the FM-FMD interface where the
phase transition is continuous ( $\kappa_{critical} = -4.084$ ). 
The figures provide evidence for the
existence of the FM-FMD phase transition at a reasonably large
gauge coupling.}
\label{vat2p0}
\end{figure}

\begin{figure}[!p]
\centering
\subfigure{
\includegraphics[width=2.0in,clip]{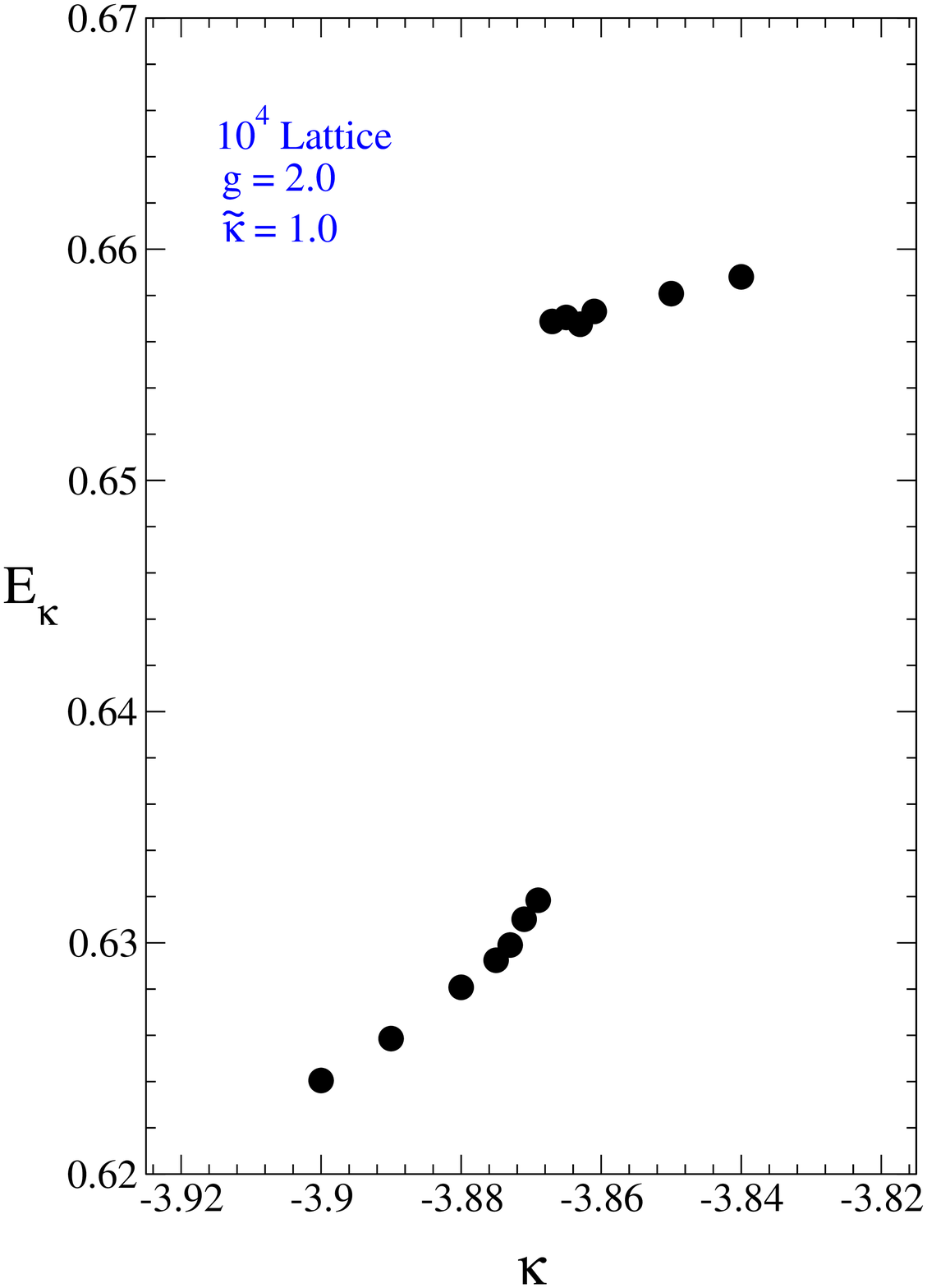}}
\subfigure{
\includegraphics[width=2.0in,clip]{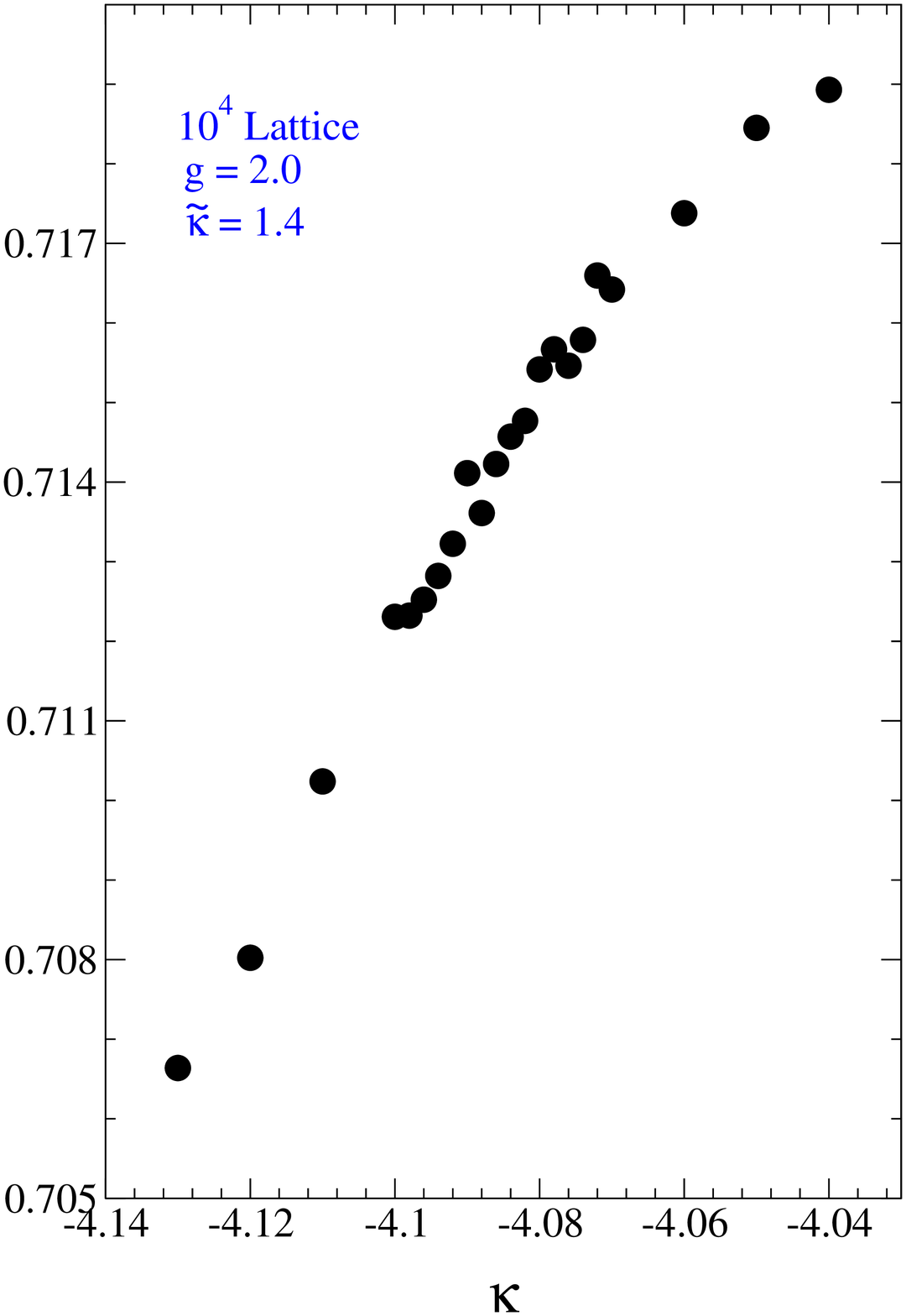}}
\caption{$E_{\kappa}$ as a function of $\kappa$
across the critical point at a large value of the gauge coupling
(left) in a region of the FM-FMD interface where the phase transition
is first-order ( $\kappa_{critical} = -3.868$ ) and (right) in a region 
of the FM-FMD interface where the
phase transition is continuous ( $\kappa_{critical} = -4.084$ ). 
The figures provide evidence for the
existence of the tricritical point at a reasonably large
gauge coupling.}
\label{ekat2p0}
\end{figure}

\begin{figure}[!t]
\centering
\subfigure{
\includegraphics[width=3in,clip]{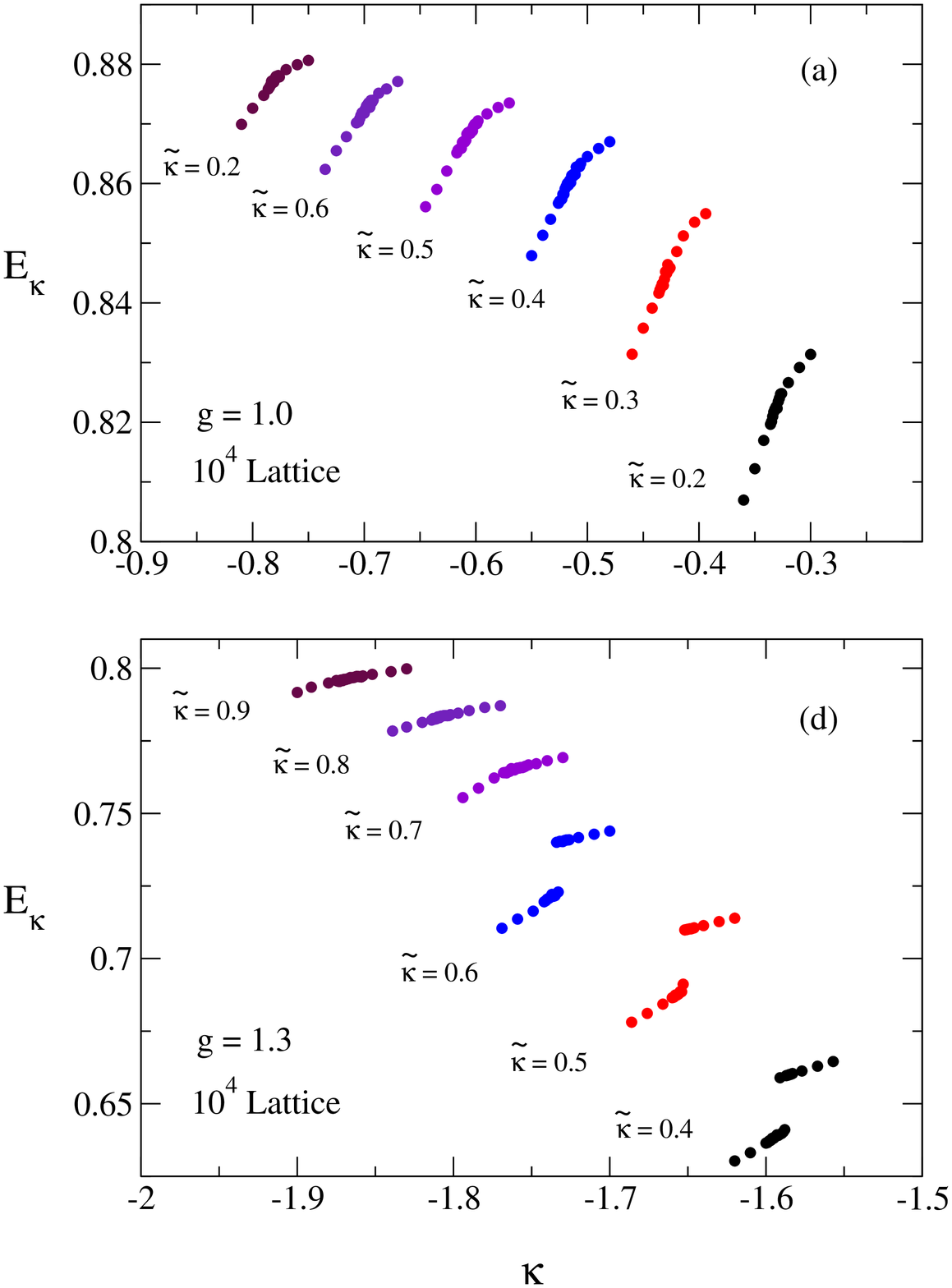}}
\subfigure{
\includegraphics[width=3in,clip]{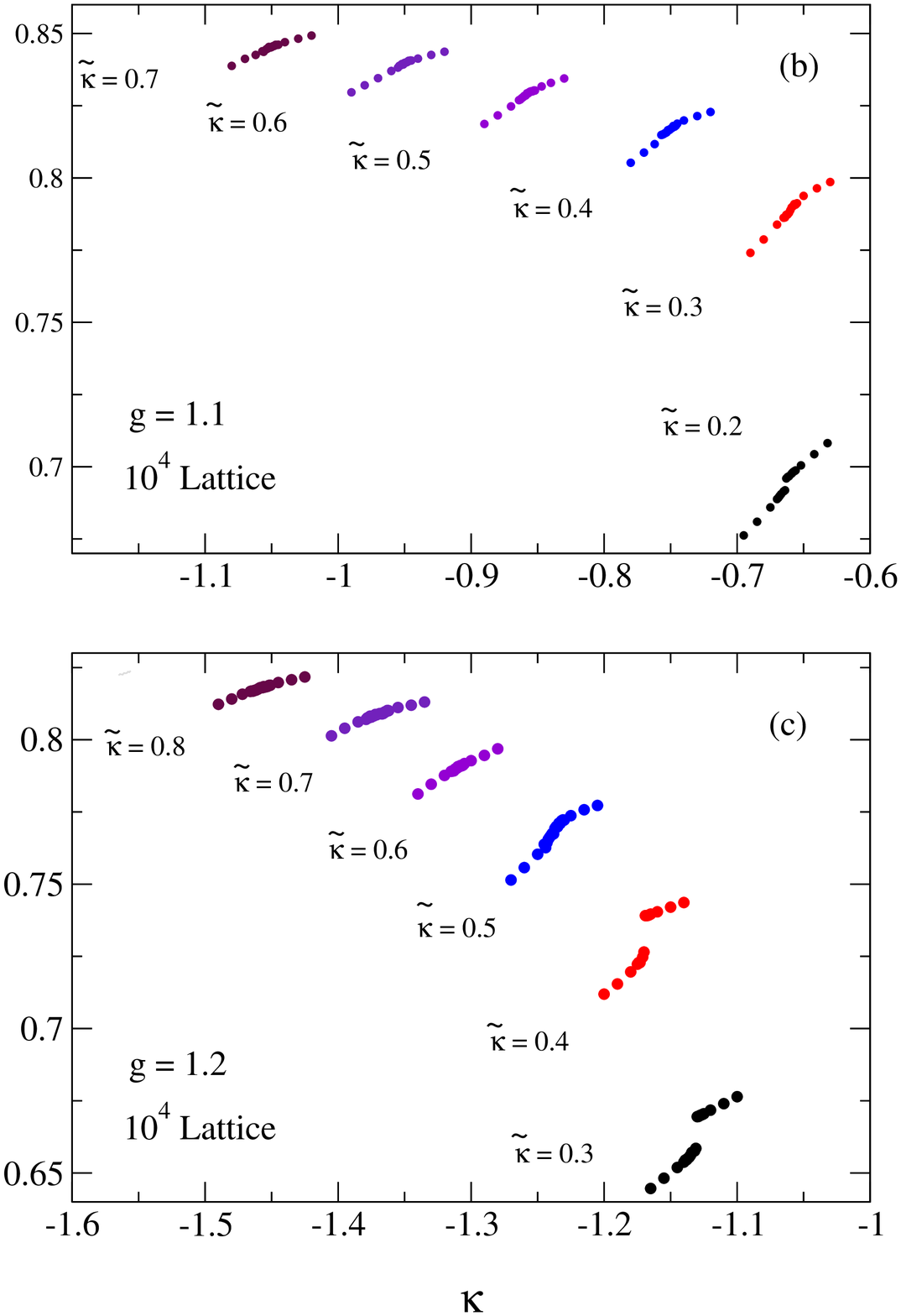}}
\caption{$E_{\kappa}$ as a function of $\kappa$ across the critical points
for a range of values of $\tilde{\kappa}$ along the FM-FMD phase transition
at four different gauge couplings: (a) g=1.0, (b) g=1.1, (c) g=1.2,
and (d) g=1.3. The value of the couplings at which the discrete jump in
$E_{\kappa}$ disappear is identified as the tricritical point.}
\label{trcrpt}
\end{figure}

\begin{figure}[!t]
\centering
\subfigure{
\includegraphics[width=2.0in,clip]{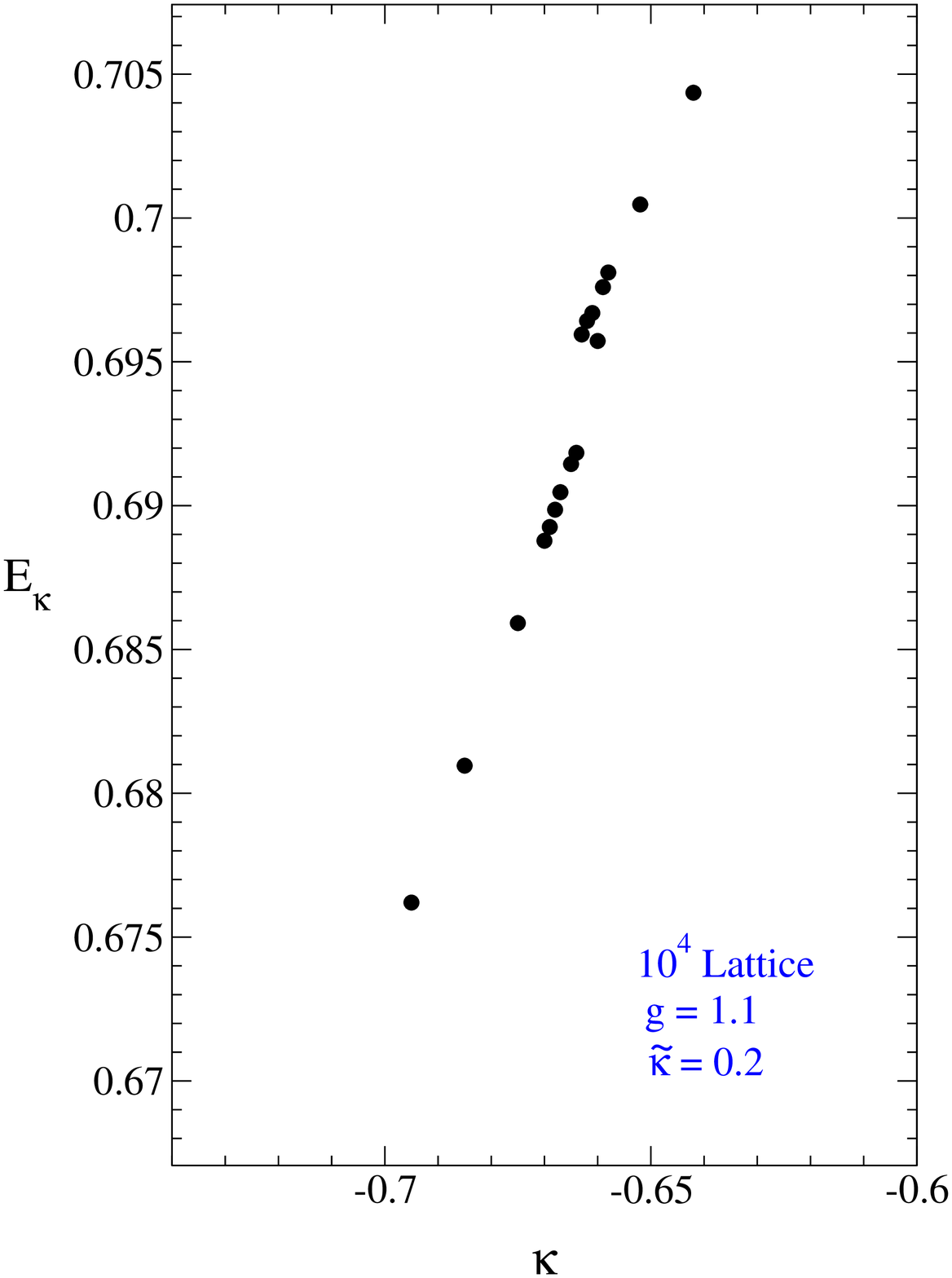}}
\subfigure{
\includegraphics[width=2.0in,clip]{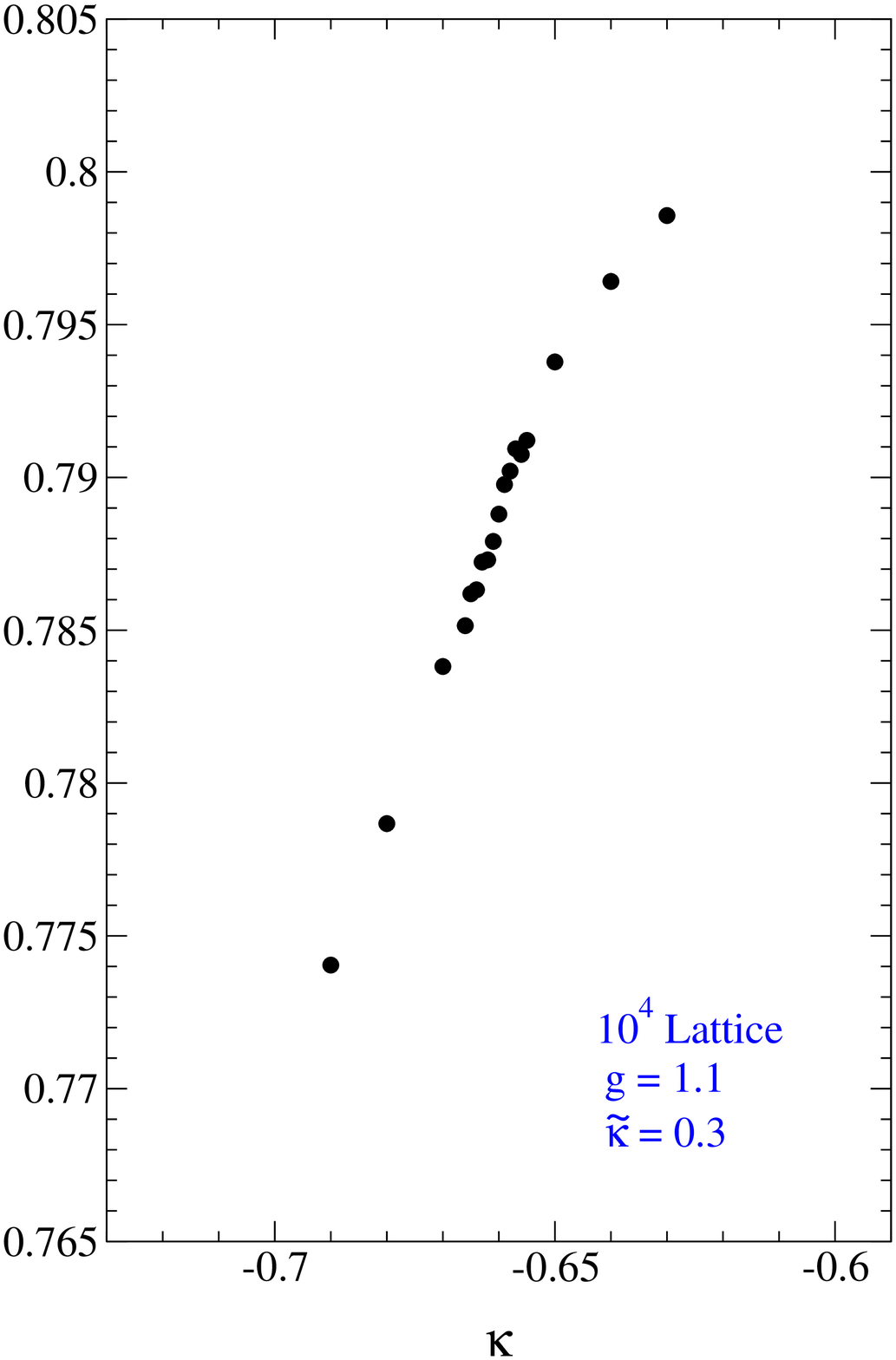}}
\caption{Blow up of $E_{\kappa}$ vs $\kappa$ plot at $g = 1.1$ and
$\tilde{\kappa}= 0.2$ and $0.3$ to show the presence of the minute discrete
jump in $E_{\kappa}$ and its disappearance. The critical values of 
$\kappa$ at $\tilde{\kappa} = 0.2$ is -0.664 and 
at $\tilde{\kappa} = 0.3$ is -0.659.}
\label{blowups}
\end{figure}

\begin{figure}[!t]
\centering
\subfigure{
\includegraphics[width=3in,clip]{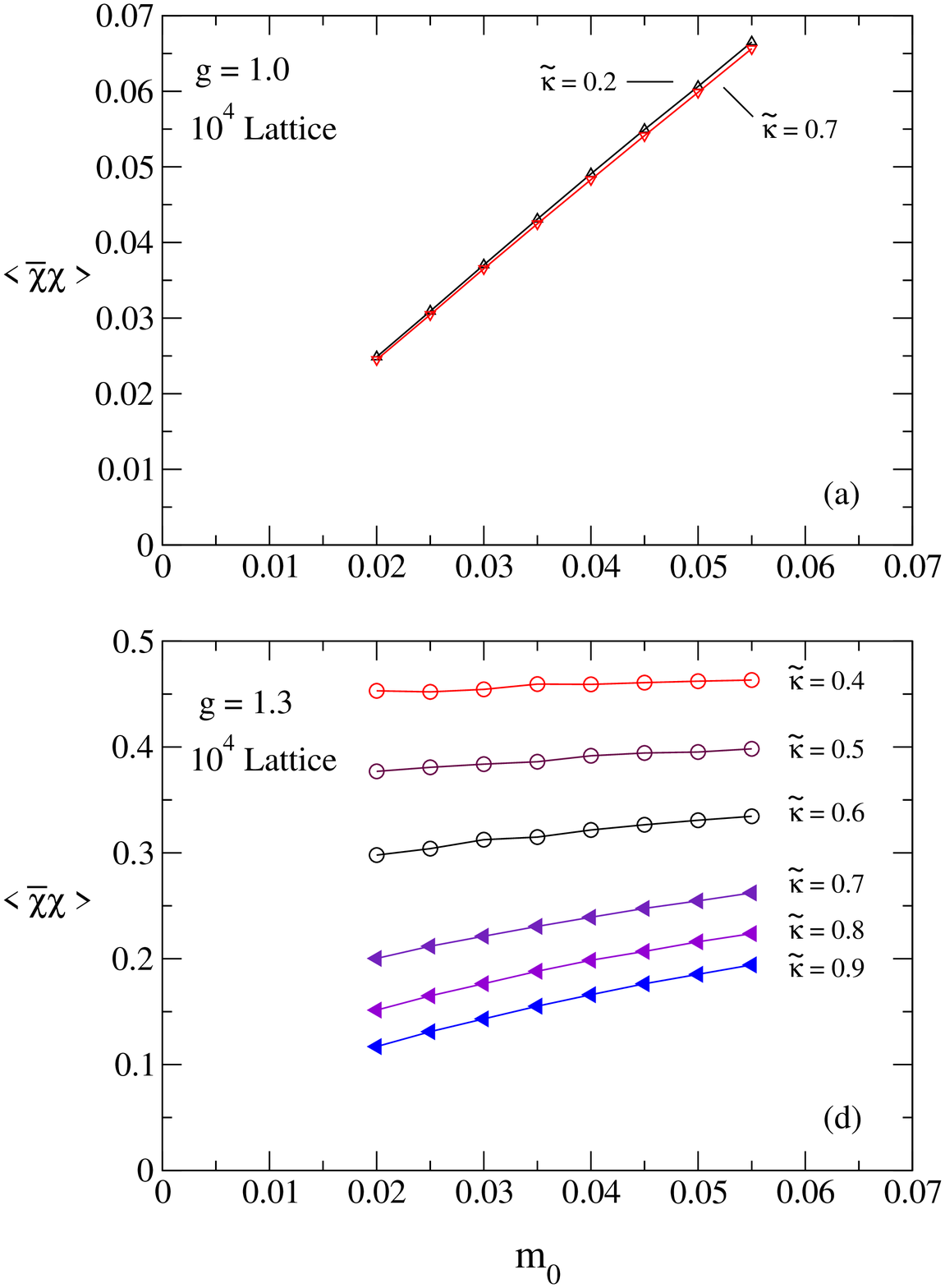}}
\subfigure{
\includegraphics[width=3in,clip]{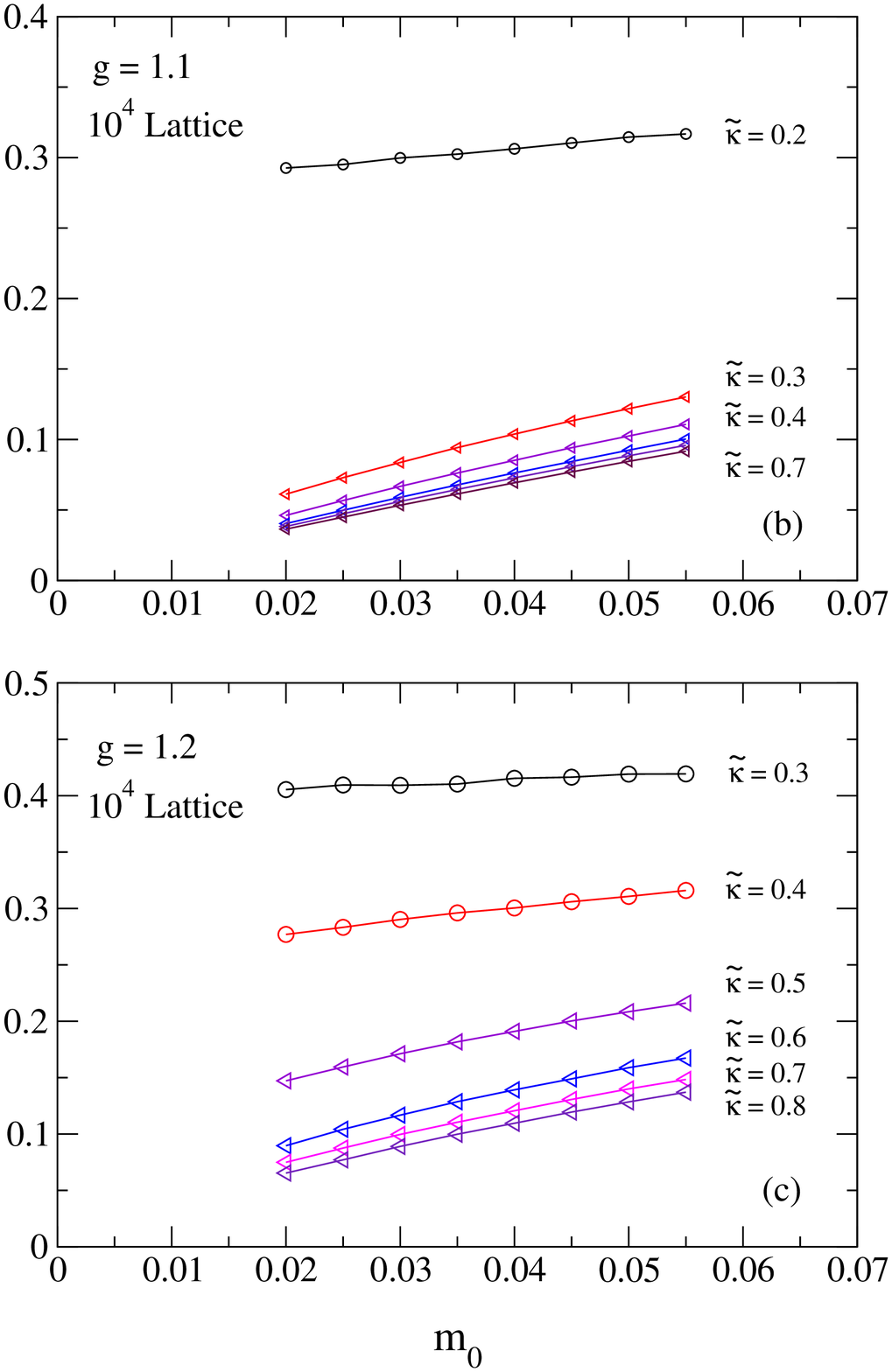}}
\caption{Quenched chiral condensates $\langle\bar{\chi}\chi\rangle$
as a function of the bare fermionic mass $m_0$
for a range of values of $\tilde{\kappa}$ along the FM-FMD phase
transition at four different gauge couplings: (a) g=1.0, (b) g=1.1,
(c) g=1.2, and (d) g=1.3. The values of $\kappa$ have been appropriately
chosen to stay close to the critical points. 
The figures demonstrate the existence of a chiral
phase transition and indicate the position where it intersects with the
FM-FMD interface.}
\label{chitr}
\end{figure}

\begin{figure}[!t]
\centering
\includegraphics[width=4in,clip]{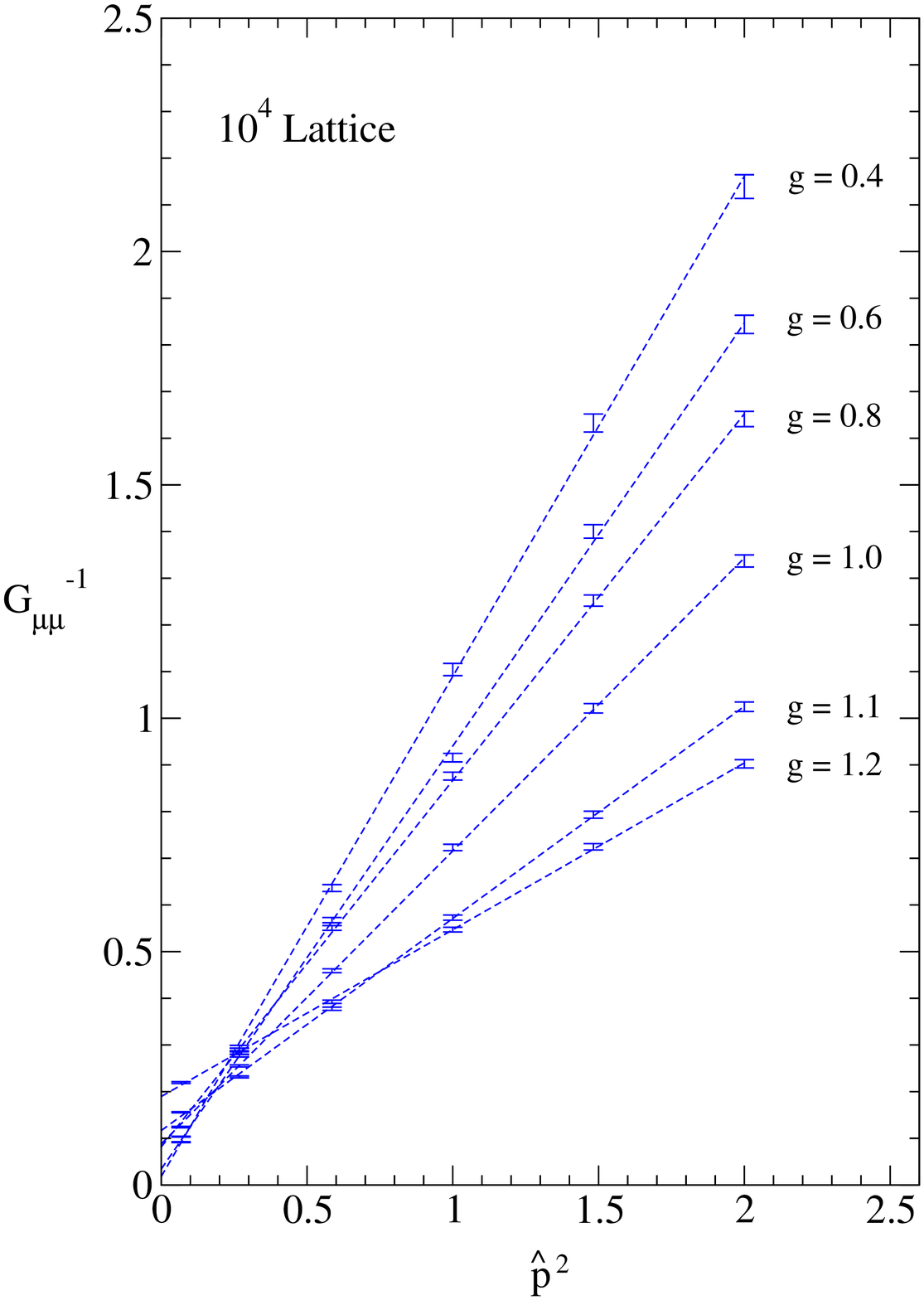}
\caption{Inverse gauge field propagator $G_{\mu\mu}^{-1}$
in momentum-space as a function of $\hat{p}^2$. A straight line
with unit slope indicates recovery of free photons.}
\label{mgprop}
\end{figure}

Since we wanted to find out the order of the phase transition, which
is typically inferred from the continuity or discontinuity of some
observable at the critical point, the critical points needed to be determined 
with extreme care and precision in this investigation.  

In \cite{paper1}, to obtain the phase diagram of the 
gauge-fixed pure $U(1)$ theory,
given by the action (\ref{action}), in $(\kappa,\,\tilde{\kappa})$
-plane for fixed values of the gauge coupling $g$, we defined the following
observables (for a $L^4$-lattice):
\bea
E_P &=& \frac{1}{6L^4} \left\langle \sum_{x, \mu<\nu} {\rm Re}\,
U_{\mu \nu x} \right\rangle \label{obsep} \\
E_\kappa &=& \frac{1}{4L^4} \left\langle \sum_{x, \mu} {\rm Re}\,
U_{\mu x} \right\rangle \label{obsek} \\
V &=& \left\langle \sqrt{\frac{1}{4} \sum_\mu \left(\frac{1}{L^4}
\sum_x {\rm Im}\, U_{\mu x} \right)^2} \;\right\rangle. \label{obsv}
\eea
where $E_P$ and $E_\kappa$ though not order parameters, signal
phase transitions by sharp changes. We expect $E_\kappa \neq 0$
in the broken symmetric phases FM and FMD and $E_\kappa \sim 0$ 
in the symmetric (PM) phase. Besides, $E_\kappa$ is expected to be
continuous at a continuous phase transition (infinite slope in 
the infinite volume limit) and show a discrete jump at a first order    
transition \cite{bock1,bock2}. 
The true order parameter is $V$ which allows us to     
distinguish the FMD phase (where $V \neq 0$) from the other phases 
where $V \sim 0$.

In this paper, in addition to the above observables we have used the 
fluctuation $\delta V$ in the order parameter $V$ in the determination 
of the phase diagram. The fluctuation $\delta V$ 
is expected to
display a sharp peak (diverge for the infinite lattice) at the critical point. 
To show how effective this is, we compare
the variation of $V$ and $\delta V$  across the critical point in
Fig.\ref{vndelv} as an example. Fig.\ref{vndelv} clearly demonstrates 
how by looking at $\delta V$ one can unambiguously determine 
the critical point (of course for the given lattice).

Throughout this paper error bars have been omitted whenever they are
smaller than the symbol size which has been the case more often than not
in this study.

The Monte Carlo simulations were done with a 4-hit Metropolis
algorithm on a $10^4$ lattice. The autocorrelation length for all observables 
was less than 10 for $10^4$ lattices and each
expectation value was calculated from one to four thousand independent
configurations. The quality of data for each coupling was examined 
for thermalization before deciding on the number of 
independent configurations required for measurments at that coupling.

The carefully re-determined phase diagram (FM-FMD phase only), 
explored in $(\kappa, \tilde{\kappa})$-plane 
at gauge couplings $g\,=\,1.0,\,1.1,\,1.2,\,1.3$ and 2.0 is shown in 
Fig.\ref{nuphase}. There are obviously small deviations to the phase 
diagram obtained in \cite{paper1} .However there are no qualitative changes.
For each gauge coupling, the FM phase lies above the phase transition and
the FMD phase lies below it (i.e, at larger negative values of $\kappa$).

Based on our computation shown in Fig\ref{vat2p0}, in which the variation 
of $V$ with the coefficient of the counter term $\kappa$ at a reasonably large 
numerical value of the gauge coupling ($g=2.0$) is shown, 
we assert that the FM-FMD
transition will exist for arbitrarily large values of $g$ provided that the
coefficient of the counter term $\kappa$ has a sufficiently large negative
value. Moreover this phase transition is continuous if the coefficient of 
the gauge fixing term  $\tilde{\kappa}$ is adequately large 
(eg: beyond $\tilde{\kappa} \sim 1.2, g=2.0$ ). This is demonstrated in 
Fig.\ref{ekat2p0} which shows a discrete jump of $E_{\kappa}$ across the
critical point at $\tilde{\kappa}=1.0$ as opposed to a continuous change in 
$E_{\kappa}$ across the critical point at $\tilde{\kappa}=1.4$. 
This is of course, what we observed in our exploratory paper \cite{paper1} 
for  gauge couplings upto $g=1.3$ and is in complete agreement 
with what we had speculated for higher gauge couplings. In determining the 
tri-critical point, this feature has been investigated in more detail and 
care in this paper and is discussed below.

To determine the tri-critical points we looked at the variation of 
the plaquette energy $E_\kappa$ with the coefficient of the mass-counterterm 
$\kappa$ across the FM-FMD phase transition for a range of values of the 
coefficient of the gauge fixing term ${\tilde{\kappa}}$. This is shown for 
gauge couplings $g=1.0$ to $g=1.3$, in Figs.\ref{trcrpt} (a) to (d).

All the figures excepting that for $g=1.0$ exhibit the same feature: A
discrete jump of $E_\kappa$ at the critical points 
(implying a first order phase transition) below a particular value 
for ${\tilde{\kappa}}$  
and a continuous change of $E_\kappa$ across the
critical points (implying a continuous phase transition)
above the particular value of ${\tilde{\kappa}}$. 
For $g=1.0$, $E_\kappa$ is always continuous.
Where the change of $E_\kappa$ is continuous, a hint of a S-shape  
is visible, a fact which is typical of continuous phase transitions.
Please note that the critical point in each $E_\kappa-\kappa$ plot
is at the middle of the region where the data is most densely packed.

To make the minute jump in 
$E_{\kappa}$ visible at $g = 1.1, \tilde{\kappa} = 0.2$ 
and the subsequent disappearance of the discrete jump at 
$\tilde{\kappa} = 0.3$ we have provided a
blow up in Fig.\ref{blowups}. 

The figures show a clear evidence for the existence of tricritical points
which taken together forms a tricritical line cutting across the FM-FMD
surface. The tri-critical line starts above $g=1.0$. Note that at $g=1.0$
there is no FM-FMD transition for $\tilde{\kappa} \le 0.2$. This region
belongs to a different phase which has been called PM phase 
(an uninteresting phase to us which includes the longitudinal 
gauge degrees of freedom in the continuum) 
in earlier papers \cite{paper1,bock1}.

As we have argued and shown in our last paper \cite{paper1} 
the features discussed above are unlikely to be finite
size effects since the features become more pronounced as one goes to 
larger lattices. 

We have repeated the calculation for quenched chiral condensates 
with $U(1)$ charged staggered fermions near our newly determined 
critical lines. This was done to investigate the possibility 
of coincidence of the tricritical line mentioned above and the chiral phase
transition on the interface between the FM and FMD phases.

We have measured the chiral condensates
\be
\langle \bar{\chi} \chi \rangle_{m_0} = \frac{1}{L^4} \sum_x
\langle M^{-1}_{xx} \rangle \label{chcnd}
\ee
as a function of vanishing fermionic bare mass $m_0$. $M$ is the fermion 
matrix. The chiral condensates were computed
with the Gaussian noise estimator method \cite{gnem}. 
Anti-periodic boundary condition in one Euclidean direction is employed.

In Figs.\ref{chitr} (a) to (d),
for gauge couplings $g=1.0$ to $g=1.3$, 
quenched chiral condensates 
in the FM phase computed at points close
to the FM-FMD transition are 
plotted against the bare mass of staggered
fermion $m_0$, once again for a range of values of the coefficient of the
gauge fixing term $\tilde{\kappa}$, along the FM-FMD critical lines. 

Except for $g=1.0$, we see  
a clear evidence of a chiral phase transition.
There exists a critical $\tilde{\kappa}$ for a given
gauge coupling $g$ (the third parameter $\kappa$ is used to stay close to 
the transition), above which the chiral condensates tend towards zero for 
vanishing fermion mass. Moreover we note, that to our precision
the chiral phase transition on the FM-FMD interface coincides with the 
tricritical line upto $g=1.1$ (at $g=1.0$ there is no tricritical line and 
no chiral transition). However as the gauge coupling is increased the line 
of chiral phase transition on the FM-FMD surface appears to move away from
the line where the order of the FM-FMD transition changes into the 
region where the FM-FMD phase transition is continuous.

\subsection*{Nature of the continuum limit}

To probe the nature of the continuum limit at the FM-FMD transition, 
we have studied the gauge
field propagator in momentum space to see where (if anywhere) in the
critical manifold free photon propagators can be observed. The dimensionless
momentum space propagator defined in terms of the group valued gauge fields  
is

\be
G_{\mu\nu}=\frac{1}{g^2L^3T}\Big(\sum_{xy} ~Im ~U_{\mu x} ~Im ~U_{\nu y}
~\exp[ip(x-y)]\Big)
\ee
where $L$ and $T$ are the spatial and temporal extents of the lattice and
$p$ are the finite box discrete momenta. This definition obviously is 
obtained by requiring that 

\[
\lim_{a \to 0} ~a^2 ~G_{\mu\nu} \longrightarrow \frac{1}{V}\int\int d^4x d^4y 
~A_{\mu x}~A_{\mu y} ~\exp[ip(x-y)] 
\]   
where $a$ is the lattice spacing, $V$ is the volume and
$U_{\mu x} = e^{i a g A_{\mu x}}$. We have studied the momentum space
propagator $G$ as a function of
$\hat{p}^2$, $~\hat{p}_{\mu} = 2 \sin(\frac{p_\mu}{2})$ being the 
dimensionless lattice momenta,
to see if a $\frac{1}{\hat{p}^2}$ dependence can be extracted somewhere on
(near) our critical manifold. 
In a theory for {\em free photons}, this is the simplest scheme that 
can actually be legitimately used to determine the values of the bare 
couplings in our parameterization.

In our numerical simulation we have actually  
computed $G_{\mu \mu}$ for different $\mu~$ (spatial directions) and
averaged over them. A momentum  table was set up from where different input 
momenta were taken corresponding to different values of 
$\hat{p}^2= \sum_{\mu \ne \nu} {\hat{p}_\nu}^2$. Here ${\hat{p}_\nu}$ are the
{\em dimensionless} finite box discrete momenta.
In the end, the inverse of the momentum space gauge field propagator, 
$G_{\mu \mu}^{-1}$, has been plotted as a function of $\hat{p}^2$.

In order to be able to access small momenta, a $6^3~24$ anisotropic lattice 
with long temporal extent was used.  
Calculations were done on $10,000$ configurations and for the errors in 
$G^{-1}(\hat{p})$, which is a so called {\em biased} quantity, Jackknife 
was employed.

\begin{table}[h]
\begin{center}
\begin{tabular}{||c|c|c|c||}
\hline
\hline
 $~~~~~g~~~~~~$ & $~~~~~~\tilde{\kappa}~~~~~~$ & $~~~~~~\kappa~~~~~~$ &
critical $\kappa$ \\
\hline
\hline
 $0.4$ & $0.2$ & $-0.014$ & $-0.064$\\
\hline
 $0.6$ & $0.3$ & $-0.077$ & $-0.134$\\
\hline
 $0.8$ & $0.3$ & $-0.169$ & $-0.224 \leftrightarrow -0.219$\\
\hline
 $1.0$ & $0.3$ & $-0.383$ & $-0.433$\\
\hline
 $1.1$ & $0.5$ & $-0.822$ & $-0.872$\\
\hline
 $1.2$ & $0.7$ & $-1.348$ & $-1.398$\\
\hline
\end{tabular}
\end{center}
\caption{Couplings at which the momentum-space
gauge field propagator were measured on $6^3 ~24$ lattice. The
fourth column shows the nearest critical points.}
\label{mpoints}
\end{table}

Fig.\ref{mgprop} shows $G_{\mu \mu}^{-1}$ as a function of $\hat{p}^2$ for
gauge couplings $g=0.4, ~0.6, ~0.8, ~1.0, ~1.1, ~1.2$. 
The measurments have
been made close to the FM-FMD phase transition from within the FM phase.
The values of the critical couplings and the points at which the measurement
has been made is listed in Table \ref{mpoints}. 
The figure clearly shows that the
dependence of $G_{\mu \mu}^{-1}$ on $\hat{p}^2$ is linear for all gauge
couplings.   
However as the gauge coupling grows larger the slope of the straight lines
are seen to diminish. Our desired value for the slope $=1$ is obtained 
between $g=0.4$ and $g=0.6$. So this is where we claim to have recovered
free photons. Moreover around $g=0.4$ the intercept on the 
$G_{\mu \mu}^{-1}$ axis is vanishingly small already on
the lattice, indicating that we have actually recovered {\em massless} free
photons. Our results are consistent with an earlier calculation done
at $g=0.4$ \cite{bock1}.    


\vspace{-0.3cm}

\section*{Conclusions and future}

Our study with the particular regularization of compact U(1) pure gauge
theory with an extended parameter space has confirmed that the speculations
made in our last paper \cite{paper1} were largely correct: We have shown
that there is clearly a continuum limit at arbitrarily large gauge 
couplings provided that the coupling associated with the coefficient of the 
gauge fixing term is kept adequately large and the coefficient of the mass
counter term is sufficiently negative.

We have located the line on the FM-FMD phase transition where its order
changes from first to continuous. We have also located the line on the
FM-FMD surface where the chiral phase transition intersects. As for the 
coincidence of these two lines our results appear to have deviated slightly
from our speculation. While at smaller gauge coupling (e.g $g=1.0$) 
the lines clearly coincide, as we move to higher gauge couplings the chiral
phase transition seems to move into the region where the FM-FMD
transition is continuous (e.g $g=1.3$). Significantly therefore, at large 
gauge couplings the chiral phase transition on the FM-FMD interface continues
to be a strong candidate for non-perturbative physics. However, it is
difficult to conclude at our precision level whether the chiral phase
transition actually moves into the continuous FM-FMD transition. Even if it
does not, the coincidence of the two transitions along the tricritical line
is very interesting from the point of view of possible nonperturbative
properties of the U(1) theory, as already pointed out in our previous paper
\cite{paper1}. 

Our momentum space propagator study has revealed that with
the continuum limit at the FM-FMD transition,
free photons can be recovered at low gauge couplings
(in the vicinity of $g=0.4$).

In conclusion, our study of a novel regularization of compact U(1) pure
lattice gauge theory reveals that at weak bare gauge coupling, as expected,
free photons emerge, while at larger bare couplings the interference of a
chiral transition (in this case obtained with U(1) charged quenched
fermions) may lead to a continuum limit with interesting
nonperturbative properties. With introduction of fermions, the issue of 
triviality is open for investigation on all continuum limits on the FM-FMD
transition.
 
\vspace{0.25in}
A major part of the numerical calculations presented in this work is carried
out  on  multiprocessor (Power4 and Power4+) IBM compute machines supported
by
the 10$^{th}$ Five Year Plan Project, Theory Division, Saha Institute of
Nuclear Physics, under the DAE, Govt. of India.

\hidestart
With the particular regularization of compact pure U(1) gauge theory with an
extended parameter space, we have shown that there is clearly
a continuum limit for the whole range of the bare gauge coupling $g$.
Evidence of a continuum limit in other regularizations of a compact
lattice U(1) gauge theory is either absent, inconclusive
\cite{azco0,jiri0,flux} 
or dependent on inclusion of fermionic interactions \cite{azco1,kogut,chiu}.

Given the long history of speculation about a confining strong coupling
U(1) gauge theory and related issues of non-triviality, we have probed the
pure gauge system by quenched staggered fermions and found a clear evidence 
for a chiral phase transition. However, the region with a nonzero chiral
condensate does not allow a continuum limit. The continuum limit in the pure
gauge theory is only attained with no chiral condensate. This is consistent
with perturbative expectations.

We have found reasonable evidence to expect that the tricritical line at
which the the order of the FM-FMD phase transition changes 
in the pure gauge theory,
coincides with the line where the chiral phase transition intersects
the FM-FMD transition. This line is the only candidate for a possible
continuum limit with nonperturbative properties like chiral condensates.  
     
The authors thank M. Golterman and P.B. Pal for reading the manuscript and
comments. This work is funded by a project under the DAE. One of the authors
(SB) acknowledges the support of the U.S. Dept. of Energy under grant  
no. DE-FG02-93ER-40762.
\hideend

\vspace{-0.3cm}


\end{document}